\documentclass[prl,showpacs,superscriptaddress,twocolumn,longbibliography]{revtex4-2}

\usepackage[utf8]{inputenc}

\usepackage{amsmath}
\usepackage{amssymb}
\usepackage{amsthm}
\usepackage{mathrsfs}
\usepackage{xcolor}
\usepackage{braket}
\usepackage{hyperref}
\usepackage{graphicx}
\usepackage{float}
\usepackage{color}

\theoremstyle{plain}

\theoremstyle{definition}

\def\W{\mathbb{W}}

\def\RR{\mathbb{R}}
\def\PP{\mathbb{P}}

\newcommand{\er}[1]{Eq.~\eqref{#1}}

\newcommand{\era}[2]{Eqs.~(\ref{#1}) and (\ref{#2})}

\begin{document}

\title{Inverse thermodynamic uncertainty relations: general upper bounds on the fluctuations of trajectory observables}

\author{George Bakewell-Smith}
\affiliation{School of Mathematical Sciences, University of Nottingham, Nottingham, NG7 2RD, United Kingdom}

\author{Federico Girotti}
\affiliation{School of Mathematical Sciences, University of Nottingham, Nottingham, NG7 2RD, United Kingdom}

\author{M\u{a}d\u{a}lin Gu\c{t}\u{a}}
\affiliation{School of Mathematical Sciences, University of Nottingham, Nottingham, NG7 2RD, United Kingdom}
\affiliation{Centre for the Mathematics and Theoretical Physics of Quantum Non-Equilibrium Systems,
University of Nottingham, Nottingham, NG7 2RD, UK}

\author{Juan P. Garrahan}
\affiliation{School of Physics and Astronomy, University of Nottingham, Nottingham, NG7 2RD, UK}
\affiliation{Centre for the Mathematics and Theoretical Physics of Quantum Non-Equilibrium Systems,
University of Nottingham, Nottingham, NG7 2RD, UK}

\begin{abstract}
Thermodynamic uncertainty relations (TURs) are general {\em lower bounds} on the size of fluctutations of dynamical observables. They have important consequences, one being that the precision of estimation of a current is limited by the amount of entropy production. Here we prove the existence of general {\em upper bounds} on the size of fluctuations of any linear combination of fluxes (including all time-integrated currents or dynamical activities) for continuous-time Markov chains. We obtain these general relations by means of concentration bound techniques. These ``inverse TURs'' are valid for all times and not only in the long time limit. We illustrate our analytical results with a simple model, and discuss  wider implications of these new relations.
\end{abstract}

\maketitle

\noindent
{\bf \em Introduction.} 
Thermodynamic uncertainty relations (TURs)
refer to a class of general {\em lower bounds} on the size of fluctuations in 
the observables of trajectories of stochastic systems. TURs were initially postulated 
as a bound on the scaled variance of time-averaged currents in the stationary state of continuous-time Markov chains \cite{Barato2015b}, and soon after proven (using ``level 2.5'' large deviation methods \cite{Maes2008,Bertini2012,Bertini2015})
to apply to the whole probability distribution \cite{Gingrich2016}.
TURs were subsequently generalised to various other dynamics and observables, including for finite times \cite{Pietzonka2017,Horowitz2017}, for 
discrete-time Markov dynamics \cite{Proesmans2017}, for the fluctuations of first-passage times \cite{Garrahan2017,Gingrich2017} and for open quantum systems~\cite{Brandner2018,Carollo2019,Guarnieri2019,Hasegawa2020}, among many other extensions and alternative derivations (see for example \cite{Polettini2016,Dechant2018,Barato2018,Nardini2018,Chun2019,Koyuk2019,Koyuk2020,Fischer2020,Dechant2020,Liu2020}). For a review see \cite{Horowitz2020}.

The most widely considered form of the TUR is for the relative uncertainty (variance over mean squared) of a time-integrated current being larger than (twice) the inverse of the entropy production. This has immediate dual consequences for inference and estimation \cite{Barato2015b,Horowitz2020}: increased precision in the estimation of the value of a current from a stochastic trajectory requires increasing the dissipation, or alternatively, the value of the entropy production can be inferred from the fluctuations of one or more specific currents which might be easier to access. Similar uses of the TUR can be formulated using the dynamical activity \cite{Garrahan2007,Lecomte2007,Maes2020} for the estimation of time-symmetric observables \cite{Garrahan2017,Horowitz2020}.

Despite their success and generality, a limitation of TURs is that they only provide lower bounds on the size of fluctuations: except in the few cases where they are tight, inference on the observable of interest is hindered by the absence of a corresponding upper bound. Here we correct this issue by introducing a class of {\em general upper bounds}
for fluctuations of trajectory observables
consisting of linear combination of fluxes 
(number of jumps between configurations \cite{Maes2008}) of a continuous-time Markov chain, which includes all currents and activities. For lack of a better name, we call these ``inverse thermodynamic uncertainty relations''. The inverse TURs are valid for all times, and like the large deviation formulation of the TURs, they bound fluctuations at all levels. We prove these general relations using concentration bound techniques \cite{Lezaud1998,Glynn2002,Jiang2018,Fan2021,Benoist2021,Girotti2022}. 

\smallskip 

\noindent
{\bf \em Notation and definitions.} Let ${\bf X}:=(X_t)_{t \geq 0}$ be a continuous-time Markov chain taking values in the finite state space $E$ with generator $\W = \sum_{x \neq y} w_{xy} |x \rangle \langle y| - \sum_x w_{xx} |x \rangle \langle x|$, with $x,y \in E$. If $X_0$ is distributed according to some probability measure $\nu$ on the state space, we denote by $\PP_\nu$ the law of ${\bf X}$ and we use $\mathbb{E}_\nu$ for the corresponding expected value. We assume that ${\bf X}$ is irreducible with unique invariant measure (i.e.\ stationary state) $\pi$. We are interested in studying fluctuations of observables of the trajectory ${\bf X}$
of the form
\[A(t)=\sum_{x \neq y} a_{xy}N_{xy}(t),
\]
where $a_{xy}$ are arbitrary real numbers with $\sum |a_{xy}| >0$, and $N_{xy}(t)$ are the elementary {\em fluxes}, that is, the number of jumps from $x$ to $y$ up to time $t$ in ${\bf X}$. For a time-integrated current the coefficients are antisymmetric, while for counting observables (such as the activity), they are symmetric. 

The fluctuations of $A(t)$ in the long time satisfy the following theorems \cite{Touchette2009}: 
\\
(i) Strong Law of Large Numbers (holds almost surely)
    \[\lim_{t \rightarrow +\infty} \frac{A(t)}{t}=\langle a \rangle_\pi:=\sum_{x \neq y} \pi_x w_{xy}a_{xy} .
    \] 
(iii) Central Limit Theorem (small deviations; holds in distribution)
    \[\lim_{t \rightarrow +\infty} \frac{A(t)-t\langle a \rangle_\pi}{\sqrt{t}} = {\cal N}(0,\sigma_\infty^2)
    \]
    where $\sigma^2_\infty=\lim_{t \rightarrow +\infty}\sigma_\nu^2(t)/t$ and $\sigma_\nu^2(t)$ is the variance of $A(t)$ if $\nu$ is the initial distribution (notice, however, that the limit does not depend on $\nu$).
\\
 (iii) Large Deviation Principle
    \[\PP_\nu \left ( \frac{A(t)}{t} = \langle a \rangle_\pi+ \Delta a\right ) \asymp e^{-t I(\Delta a)}
    \;\;\; 
    \text{for every } \Delta a \in \RR
    \]
    for some rate function $I:\RR \rightarrow [0,+\infty]$ which in general is hard to determine and admits an explicit analytic expression only for particular models.

In order to state our main result below we need to introduce the following quantities. In the stationary state $\pi$, the average of $A$ per unit time is $\langle a \rangle_\pi=\sum_{x \neq y} \pi_xw_{xy} a_{xy}$, while its {\em static approximate variance} is $\langle a^2 \rangle_\pi$, with $\langle a^2 \rangle_\pi=\sum_{x \neq y} \pi_xw_{xy} a^2_{xy}$ (it is the variance of the random variable $\sum_{x \neq y}a_{xy}\tilde{N}_{xy}$, where we approximate $N_{xy}(t)/t$ with independent Poisson random variables $\tilde{N}_{xy}$ with intensity $\pi_xw_{xy}$). The maximum escape rate is $q=\max_{x} w_{xx}$, and $c=\max_{x\neq y} |a_{xy}|$ the maximum amplitude of the coefficients that define the observable. Since we do not assume that $\W$ is reversible, we denote by $\varepsilon$ the spectral gap of the symmetrization $\Re(\W)=(\W+\W^\dagger)/2$, where the adjoint is taken with respect to the inner product induced by the stationary state $\pi$. Finally, the average dynamical activity per unit of time at stationarity is $\langle k \rangle_\pi:=\sum_{x \neq y} \pi_x w_{xy}$.

\smallskip 

\noindent
{\bf \em Main results.} We now state our three main results:
\\
(R1) The variance $\sigma_\pi^2(t)$ of any time-integrated current or flux observable $A(t)$ in the stationary state has the general upper bound
    \begin{equation}
        \sigma_\pi^2(t) \leq t \langle a^2 \rangle_\pi\left (1 + \frac{2q}{\varepsilon} \right )
        \label{R1}            
    \end{equation}   
Note that this is valid for trajectories of any length $t$.
\\
(R2) The distribution of $A(t)/t$ after time $t$ starting from an initial measure $\nu$ obeys a {\em concentration} bound
    \begin{equation} 
        \label{R2a}
        \PP_\nu \left ( \frac{A(t)}{t} \geq \langle a \rangle_\pi+\Delta a \right ) \leq C(\nu) e^{-t \tilde{I}(\Delta a)},
    \end{equation}
    where $\Delta a > 0$ is the fluctuation of $A$ away from the stationary average,
    and $C(\nu):=(\sum_{x} \nu_x^2/\pi_x)^{1/2}$ accounts for the difference between $\nu$ and the stationary  $\pi$, with $C(\pi)=1$.
    The bounding rate function is given by
    \begin{equation}
        \tilde{I}(\Delta a)=\frac{\Delta a^2}{2 \left ( \langle k \rangle_\pi c^2+\frac{2q\langle a^2 \rangle_\pi}{\varepsilon} + \frac{5cq \Delta a }{\varepsilon}\right )}
        \label{R2b}            
    \end{equation}
\\
(R3) The rate function for $A(t)/t$ is lower bounded by \er{R2b} for every $\Delta a > 0 $, that is
    \begin{equation}    
        \tilde{I}(\Delta a) \leq 
        I(\Delta a) .
        \label{R3}
    \end{equation}
(R1)-(R3) are extended straightforwardly to $\Delta a < 0$ by considering the observable $-A(t)$.

\begin{figure}[ht!]
    \begin{center}
    \includegraphics[width=0.9\linewidth]{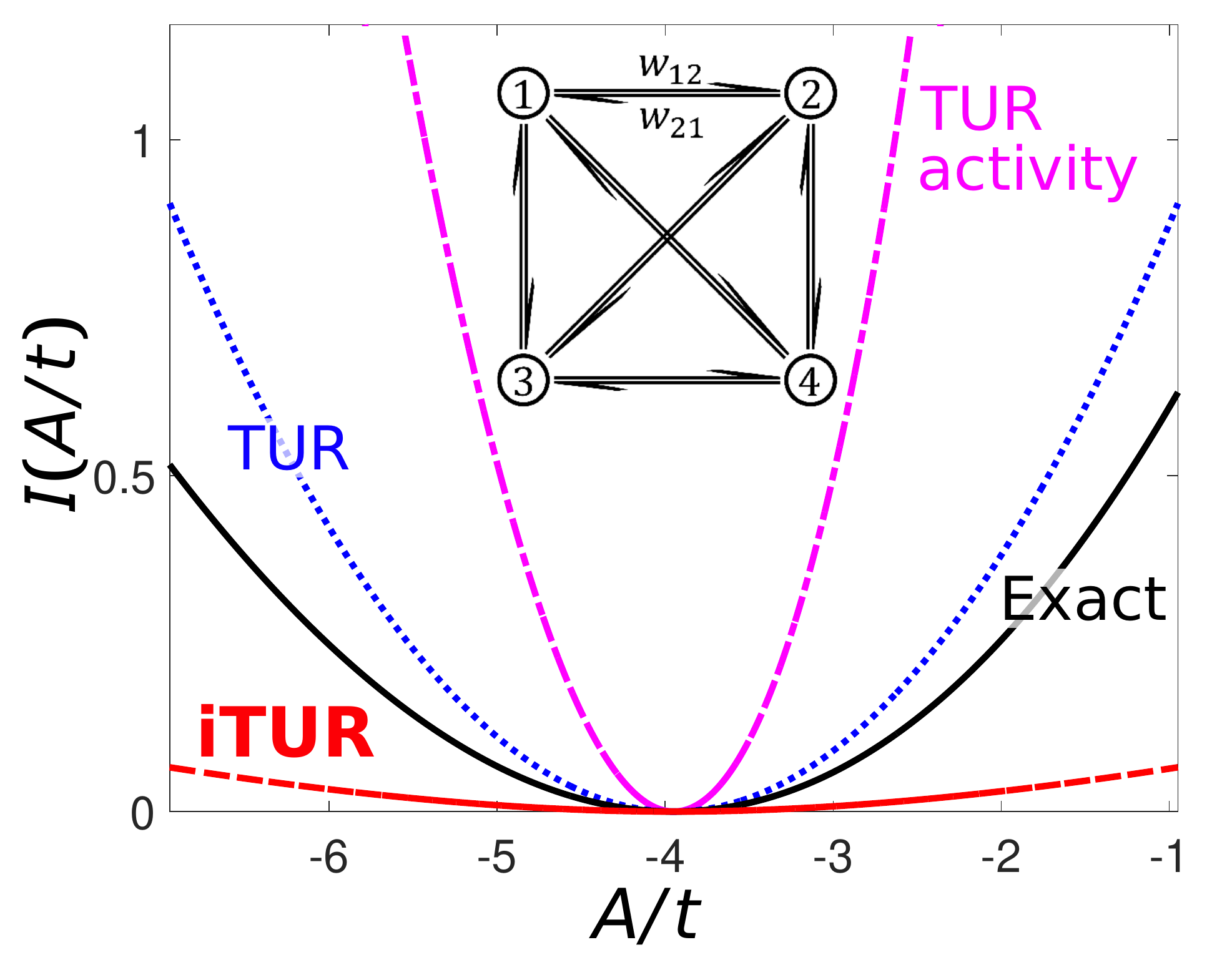}
    \end{center}
    \caption{{\bf Upper bound on current fluctuations.}  
    The full (black) curve shows the exact rate function $I(A/t)$ for the current defined by $a_{12} = 0.9, a_{13} = -0.9, a_{14} = -0.9, a_{23} = 0.9, a_{24} = -0.9, a_{34} = 0.9$. The rate function is upper bounded by the TURs: the dotted (blue) curve is the standard TUR using the entropy production, while the dot-dashed (pink) is the TUR with the dynamical activity. 
    The dashed (red) curve is the inverse TUR: it lowers bound the rate function, which corresponds to an upper bound to fluctuations of the current at all orders. 
    Inset: sketch of the 4-state model.
    }
    \label{rf_pic}
\end{figure}

\begin{figure*}[ht!]
    \begin{center}
    \includegraphics[width=\linewidth]{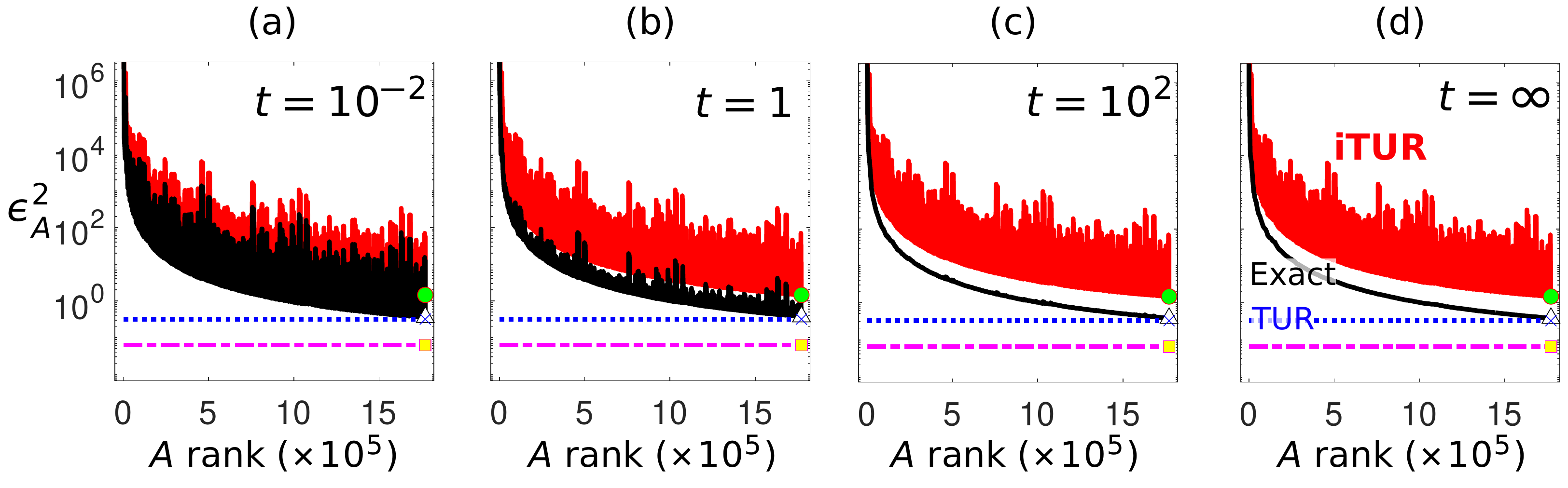}
    \end{center}
    \caption{{\bf Lower and upper bounds on the estimation error.} 
    (a) Relative error $\epsilon_A^2$ 
    for estimating a current $A$ from a trajectory of length $t=10^{-2}$ in the NESS of the model of Fig.~\ref{rf_pic}. We show results for $20^6$ different currents $A  \in {\mathbb T}$. 
    The full (black) curve is the exact error. The standard TUR, dotted (blue) line, and the activity TUR, dot-dashed (pink) line, 
    provide lower bounds to the error which are independent of $A$. 
    The inverse TUR, dashed (red) curve, gives an upper bound to the error which varies with $A$. 
    (b-d) Same for times $t=1, 10^2, \infty$, respectively. 
    The data in all panels is ranked according to decreasing values of the error at $t=\infty$. 
    For comparison, the $A$ corresponding to entropy production  is shown according to the same ranking: TUR bound (green circle), exact (white triangle), TUR (blue cross), activity TUR (yellow square).}
    \label{pres_comp}
\end{figure*}

\smallskip 
\noindent
{\bf \em Inverse TURs and bounds on precision.} 
The most direct use of TURs is in bounding the precision for estimating a current from its time-average over a trajectory in a non-equilibrium stationary state (NESS) $\pi$. We define the relative error $\epsilon_A$ of observable $A$ as the ratio between the variance of $A$ and its average squared multiplied by $t$ 
\begin{align}
    \epsilon_A^2 
    = 
    t \frac{\sigma_\pi^2(t)}{\langle A \rangle_\pi^2} =
    \frac{\sigma_\pi^2(t)}{t \langle a \rangle_\pi^2}.
    \label{err}
\end{align}
From the standard application of the TUR together with the ``inverse TUR'' \er{R1} we can bound the relative error from {\em below and above}
\begin{align}
    \frac{2}{\Sigma_\pi} 
    \leq 
    \epsilon_A^2 
    \leq 
    \frac{\langle a^2 \rangle_\pi}
    {\langle a \rangle_\pi^2}
    \left(1 + \frac{2q}{\varepsilon} \right)
    \label{lowerupper}
\end{align}
where $\Sigma_\pi = \sum_{x \neq y} \pi_x w_{xy} \log(\pi_x w_{xy}/\pi_y w_{yx})$
is the average entropy production rate in the NESS. The lower bound is the known statement that a smaller error requires larger dissipation. The upper bound 
states that the error is controlled by the static uncertainty 
and by the ratio between the largest and smallest relaxation rates in the dynamics. Note that $2 q/\varepsilon \geq 1$ (see \cite[Lemma 1]{SM}).

As an illustration of \era{R3}{lowerupper} we consider the fluctuations of currents in the 4-state model of Ref.~\cite{Gingrich2016}. The network of elementary transitions is shown in the inset of Fig.~\ref{rf_pic}. The rates are as in Ref.~\cite{Gingrich2016}, $w_{12} = 3$, $w_{13} = 10$, $w_{14} = 9$, $w_{21} = 10$, $w_{23} = 1$, $w_{24} = 2$, $w_{31} = 6$, $w_{32} =4$, $w_{34} = 1$, $w_{41} = 7$, $w_{42} = 9$ and $w_{43} = 5$. A current is defined by the values of the six coefficients 
$a_{x > y} = - a_{x<y}$
which we take in the range $a_{xy}\in [-1,1]$. 
To perform the analysis, we construct a mesh across the space of current observables $\mathbb{T}=[-1,1]^6$. We discretise with a grid spacing of $10^{-1}$. Each point on this six-dimensional mesh corresponds to a different current, and in this way we get a reasonable representation of $\mathbb{T}$. 

Figure~\ref{rf_pic} shows the bounds for the long-time limit rate function $I(A/t)$ for one such current $A \in \mathbb{T}$. The full (black) curve is the exact rate function.
It is calculated from the ``tilted'' generator 
$\W_u = \sum_{x \neq y} e^{u a_{xy}} w_{xy} |x \rangle\langle y | - \sum_x w_{xx} |x \rangle \langle x|$ as follows \cite{Touchette2009}:
(i) the moment generating function (MGF) of $A$ is $Z_{\pi,t}(u) := \mathbb{E}_\pi[e^{uA(t)}]=\langle \pi | e^{t \W_u} | - \rangle$, where $| - \rangle = \sum_x | x \rangle$ is the ``flat state''; (ii) at long times $Z_{\pi,t}(u) \asymp e^{t \Lambda(u)}$, where the scaled cumulant generating function (SCGF) $\Lambda(u)$ is the largest eigenvalue of $\W_u$; (iii) the rate function is obtained via a Legendre transform, $I(a) = \sup_u [u a - \Lambda(u)]$.

The dotted (blue) curve in Fig.~\ref{rf_pic} is the usual TUR using the entropy production
\cite{Gingrich2016}. The dot-dashed (pink) curve is the alternative TUR which instead of $\Sigma_\pi$ uses the average dynamical activity, $\langle k \rangle_\pi = \sum_{x \neq y} \pi_x w_{xy}$ \cite{Garrahan2017,Di-Terlizzi2018}. Both these curves are above the true rate function, thus providing the usual lower bounds on the size of the fluctuations of $A$. The dashed (red) curve is \er{R3} and lies below the rate function: this inverse TUR is an upper bound on the size of fluctuations of $A$ at all orders. 

Figure \ref{pres_comp} shows the bounds \eqref{lowerupper} on the precision error \eqref{err}, for all currents in the grid that scans $\mathbb{T}$, both at finite and infinite $t$. 
The full (black) curves are the exact error $\epsilon_A^2$, where the first two moments of $A$ are obtained from the first and second derivatives of the MGF $Z_{\pi,t}(u)$ evaluated at $u=0$. The errors for all the currents are plotted with rank ordered by their value at $t = \infty$, so that in Fig.~\ref{pres_comp}(d) they are monotonically decreasing. The dotted (blue) lines are the lower bounds from the TUR at either finite \cite{Pietzonka2017} or infinite \cite{Barato2015b} times. The dot-dashed (pink) lines are the activity TUR, where in the l.h.s.\ of \er{lowerupper} $\Sigma_\pi$ is replaced by $2\langle k \rangle_\pi$. 
As the TURs do not depend on the details of the current that they bound, these curves appear constant in Fig.~\ref{pres_comp}.

Figure \ref{pres_comp} also shows the inverse TUR from the r.h.s.\ of \er{lowerupper} as full (red) curves. This gives an upper bound to the error. As the inverse TUR contains information about the specific current through its static average and second moment, the bound tracks the change in shape of the exact error. Furthermore, in many instances the ratio of the relative value of the 
upper bound to the error is smaller than that of the error to the lower bound (note the log scale in the plots).

\smallskip 
\noindent
{\bf \em Derivation of results.} We now give the main steps for the proofs of results R1-R3. For details see \cite{SM}. 

Result R3 follows easily from R2: indeed
from the definition of large deviation principle and \er{R2a} is easy to see that
\begin{align}
    \inf_{\Delta a^\prime>\Delta a} I(\Delta a^\prime) 
    &
    \geq \underset{t \rightarrow +\infty}{\rm liminf} -\frac{1}{t} \log \PP_\nu  \left ( \frac{A(t)}{t} 
    \geq \langle a \rangle_\pi+\Delta a\right )  
    \nonumber
    \\
    &
    \geq \tilde{I}(\Delta a).
\end{align}
The inequality holds for $\Delta a^\prime \geq \Delta a$ due to the continuity of $\tilde{I}$ and $\inf_{\Delta a^\prime \geq \Delta a}I(\Delta a^\prime)=I(\Delta a)$ because it is non-decreasing for $\Delta a \geq 0$.

To obtain R1 and R2, the first step is upper bounding the moment generating function of $A(t)$: we show that for every $u \geq 0$ the following holds true
\begin{equation} 
    \label{eq:mgfbound}
     Z_{\nu,t}(u)\leq C(\nu)e^{t \tilde{\Lambda}(u)},
\end{equation}
where
\begin{align}
    \tilde{\Lambda}(u) 
    = 
    \sum_{x \neq y} 
    &
    \pi_x w_{xy}(e^{u a_{xy}}-1) 
    \nonumber \\
    & 
    + \frac{q \langle a^2\rangle_\pi u^2}{\varepsilon\left (1-\frac{5qcu}{\varepsilon} \right)}
    \label{eq:Lambda.tilde}
\end{align}
if $0 \leq u < \frac{\varepsilon}{5qc}$ and $+\infty$ otherwise. The bound \eqref{eq:mgfbound} consists of two parts: the first summation is the SCGF of $\sum_{x \neq y} a_{xy} \tilde{N}_{xy}$, where $\tilde{N}_{xy}$ are independent Poisson random variables with rates $\pi_xw_{xy}$; the second term 
takes care of the correlations between the jumps of the Markov chain. Equations \eqref{eq:mgfbound} and \eqref{eq:Lambda.tilde} imply that for non-negative $u \in \RR$
\begin{align}
    &\log(Z_{\pi,t}(u))=
    \langle a \rangle_\pi t u + \frac{1}{2} \sigma_\pi^2(t) u^2 + o(u^2) 
    \nonumber \\
    &\leq 
    \langle a \rangle_\pi t u + \frac{1}{2} t\langle a^2 \rangle_\pi\left (1 + \frac{2q}{\varepsilon} \right ) u^2 + o(u^2)=\tilde{\Lambda}(u),
\end{align}
and R1 follows easily. In order to show R2, first we upper bound $\tilde{\Lambda}$ with a different function $\tilde{\tilde{\Lambda}}$ using the fact that $\sum_{x \neq y} a_{xy} \tilde{N}_{xy} \leq c \sum_{x \neq y}  \tilde{N}_{xy}$ and that $\sum_{x \neq y}  \tilde{N}_{xy}$ is distributed as a Poisson random variable with intensity given by the average dynamical activity rate $\langle k \rangle_\pi$. Applying Chernoff's method (\cite[Section 2.2]{BLM13}), we get that
\begin{align}
    \PP_\nu 
    &
    \left ( \frac{A(t)}{t} \geq \langle a \rangle_\pi+\Delta a\right ) 
    \nonumber \\
    &\leq 
    C(\nu)e^{-t\sup_{u \geq 0}[u (\langle a \rangle_\pi + \Delta a) -\tilde{\tilde{\Lambda}}(u)]} 
\end{align}
and to obtain R2 we prove (\cite{SM}) that $\tilde{I}$ in \eqref{R2b} is dominated by the Fenchel-Legendre transform of $\tilde{\tilde{\Lambda}}$, that is 
\begin{equation}
    \tilde{I}(\Delta a) \leq \sup_{u \geq 0} [ (\langle a \rangle_\pi + \Delta a)u-\tilde{\tilde{\Lambda}}(u) ]
\end{equation}

The technical part consists in proving  \er{eq:mgfbound} see \cite{SM} for details. As we already mentioned, a simple calculation shows that 
\begin{equation}
    Z_{\nu,t}(u)=
    \bra{\nu}e^{t\W_u} \ket{-}, \quad u \in \RR
\end{equation}
for the ``tilted'' generator $\W_u = \sum_{x \neq y} (e^{u a_{xy}}-1) w_{xy} |x \rangle \langle y| + \W$, which is an analytic perturbation of $\W$. We consider the action of $\W_u$ on the inner product space of complex functions defined on the state space $E$ endowed with the inner product
\[\langle h,f \rangle_\pi=\sum_{x} \pi_x \bar{h}_x f_x.
\]
Applying Cauchy-Schwartz inequality, we obtain that 
\[Z_{\nu,t}(u) \leq C(\nu) \|e^{t \W_u}\|,\]
where the norm of $e^{t \W_u}$ is the one as an operator acting on the inner product space we just defined. Lumer-Phillips Theorem (\cite[Corollary 3.20, Proposition 3.23]{En01}) implies that $\|e^{t \W_u}\| \leq e^{t \lambda(u)} \leq e^{t|\lambda(u)|}$ where $\lambda(u):=\max\{z:\,z \in {\rm Sp}(\Re(\W_u))\}$. What is left to do is to upper bound $|\lambda(u)|$: for values of $u$ small enough, perturbation theory allows us to express $\lambda(u)$ as
\begin{equation}
    \lambda(u)=\sum_{x \neq y} \pi_x w_{xy} (e^{u a_{xy}}-1) + \sum_{k=2}^{+\infty} u^k \lambda^{(k)}
\end{equation}
From the explicit expression of $\lambda^{(k)}$'s we can show that
\begin{equation}
    |\lambda^{(k)}| \leq \frac{q\langle a^2 \rangle_\pi}{\varepsilon}\left (\frac{5qc}{\varepsilon}\right )^{k-2}
\end{equation}
Taking $u$ such that the geometric series converges, one gets \er{eq:mgfbound}.

\smallskip 
\noindent
{\bf \em Outlook.}
Here we have proved a general class of upper bounds on the size of fluctuations of time-integrated currents and other flux observables of trajectories. These bounds complement the lower bounds provided by the much studied thermodynamic uncertainty relations, and therefore we have tentatively called our results {\em inverse} TURs. Our bounds, results R1-R3 above, apply to fluctuations at all orders and at all times. As for TURs, the more direct use is in bounding estimation errors, see Fig.~\ref{pres_comp} for an example. We note that in contrast to the standard TUR, our bounds do depend on details of the current of interest, and in many cases they provide a much better estimate to the size of the error than the TUR. Having upper and lower bounds is important as they limit the range of values for fluctuations. 

There are many possible extensions and refinements of our results here. We have focused on continuous-time Markov chains, but analogous bounds should be obtainable for discrete time dynamics. Our bounds have as an input the spectral gap of the generator, which is not easy to obtain in systems with too large state spaces. However, the gap can be estimated from time correlations \cite{Noe2017,Bonati2021}, which could allow to extend these results to many-body systems. Further approximations may also allow to formulate the inverse TURs in terms of operationally accessible quantities. 
We also note that the classical results here will have a corresponding generalisation for open quantum dynamics by exploiting generalisations of concentration bounds to the quantum case, see e.g.~\cite{Girotti2022}. We hope to report on such extensions in future publications.

\acknowledgements
\textbf{\em Acknowledgements.-}
This work was supported by the EPSRC grant EP/T022140/1.

\bibliography{biblio}

\end{document}


\title{Inverse Thermodynamic Uncertainty Relations: General Upper Bounds on the Fluctuations of Trajectory Observables\\[4mm]
Supplemental Material}

\author{George Bakewell-Smith}
\affiliation{School of Mathematical Sciences, University of Nottingham, Nottingham, NG7 2RD, United Kingdom}

\author{Federico Girotti}
\affiliation{School of Mathematical Sciences, University of Nottingham, Nottingham, NG7 2RD, United Kingdom}

\author{M\u{a}d\u{a}lin Gu\c{t}\u{a}}
\affiliation{School of Mathematical Sciences, University of Nottingham, Nottingham, NG7 2RD, United Kingdom}
\affiliation{Centre for the Mathematics and Theoretical Physics of Quantum Non-Equilibrium Systems,
University of Nottingham, Nottingham, NG7 2RD, UK}

\author{Juan P. Garrahan}
\affiliation{School of Physics and Astronomy, University of Nottingham, Nottingham, NG7 2RD, UK}
\affiliation{Centre for the Mathematics and Theoretical Physics of Quantum Non-Equilibrium Systems,
University of Nottingham, Nottingham, NG7 2RD, UK}

\maketitle

\section{Upper bound on the Laplace transform of $A(t)$}

Recall that the the tilted generator is defined as
$$
\W_u = \sum_{x \neq y} (e^{u a_{xy}}-1) w_{xy} \ket{x} \bra{y} + \W
$$
Using Cauchy-Schwarz inequality, we can write
\[Z_{\nu,t}(u)= \left \langle \frac{\nu}{\pi} , e^{t\W_u} - \right \rangle_\pi \leq \left \|\frac{\nu}{\pi} \right \|_\pi \|e^{t\W_u}\|, \quad \|e^{t \W_u}\|:=\sup_{\|f\|_\pi=1} \|e^{t \W_u}f\|_{\pi}.
\]
Using Lumer-Phillips theorem (\cite[Corollary 3.20, Proposition 3.23]{En01}), we can further upper bound
\[\|e^{t \W_u}\|_{2} \leq e^{t\lambda(u)} \leq e^{t|\lambda(u)|}
\]
where $\lambda(u):=\max\{z:\,z \in {\rm Sp}(\Re(\W_u))\}$.

\bigskip In order to upper bound $|\lambda(u)|$, first we find an alternative expression for small $u$'s using perturbation theory. Since $\W$ is irreducible, so is $\Re(\W)$ and Perron-Frobenius theory implies that $\lambda(0)=0$ is an algebraically simple eigenvalue. Moreover, if we introduce the notation $\mathbb{J}_{xy}=w_{xy} \ket{x}\bra{y}$, we can write
\[\Re(\W_u)=\Re(\W)+\sum_{k \geq 1} \frac{u^k}{k!} \B^{(k)}, \quad \B^{(k)}:=\sum_{x \neq y} a_{xy}^k \Re(\J_{xy}).
\]
Being $\Re(\W_u)$ an analytic perturbation of $\Re(\W)$, perturbation theory (see \cite[Section 2]{Lezaud1998} and references therein for details) ensures that if we can find $\alpha, \beta >0$ such that $\|\B^{(k)}\| \leq \alpha \beta^{k-1}$ for $k \geq 1$, then for $|u| \leq (2\alpha \varepsilon^{-1}+\beta)^{-1}$ ($\varepsilon$ is the spectral gap of $\Re(\W)$) we can write
\begin{equation} \label{eq:series}\lambda(u)=\sum_{k \geq 1} u^k \lambda^{(k)}
\end{equation}
with 
\begin{equation}\label{eq:lambda.k}
\lambda^{(k)}=\sum_{p=1}^k \underbrace{\frac{(-1)^p}{p} \sum_{\substack{\nu_1+\dots +\nu_p=k,\, \nu_i\geq 1\\
\mu_1+\dots +\mu_p=p-1, \, \mu_j\geq 0}} \frac{1}{\nu_1! \cdots \nu_p!}\tr(\B^{(\nu_1)} \S^{(\mu_1)}\cdots \B^{(\nu_p)}\S^{(\mu_p)})}_{=:\lambda^{(k)}_p},
\end{equation}
where $\S^{(0)}= -\ket{-}\bra{-}$ and $\S^{(\mu)}=(\Re(\W)+\ket{-}\bra{-})^{-\mu}-\ket{-}\bra{-}$. Notice that $\|\S^{(\mu)}\|_2 \leq \varepsilon^{-\mu}$.

The following lemma provides the estimates required for bounding the convergence radius and the coefficients of the series in equation \eqref{eq:series}. We recall the following quantities:
\[q:=\max_{x \in E} w_{xx}, \quad \langle a^2\rangle_\pi:=\sum_{x \neq y} \pi_xw_{xy}a_{xy}^2, \quad c:=\max_{x \neq y} |a_{xy}|.\] 
\begin{lemma}\label{lemma:1}
The following estimates hold true for $k \geq 1$:
\begin{enumerate}
    \item $\|\B^{(k)}\| \leq qc^{k}$,
    \item $\|\B^{(k)}({\bf 1})\|_\pi\leq \langle a^2 \rangle^{1/2}_\pi q^{1/2}c^{k-1}$,
    \item $2q/\varepsilon \geq 1.$
\end{enumerate}
\end{lemma}
\begin{proof}
Notice that
\[ \left (\sum_{x \neq y}a_{xy}^k\J_{xy} \right )^\dagger=\sum_{x \neq y}\frac{\pi_y}{\pi_x}w_{yx}a_{yx}^{k} \ket{x}\bra{y}.
\]
Since $\B^{(k)}=\sum_{x \neq y} a_{xy}^k \Re(\J_{xy})$, we have that
\[\|\B^{(k)}\| \leq \left \|\sum_{x \neq y}a_{xy}^k\J_{xy} \right \|= \left \|\left (\sum_{x \neq y}a_{xy}^k\J_{xy} \right )^\dagger \right \|\]
and
\[\|\B^{(k)}\ket{-}\|_\pi \leq \max\left \{ \left \|\sum_{x \neq y}a_{xy}^k \J_{xy}\ket{-} \right \|_\pi, \left \|\sum_{x \neq y}a_{xy}^k \J_{xy}^{\dagger}\ket{-} \right \|_\pi \right \}.\]

\bigskip 1. Let $f:E \rightarrow \mathbb{R}$ be a function. Then
\[\begin{split}
    &\left \|\sum_{x \neq y}a_{xy}^k \J_{xy}(f) \right \|^2_\pi=\sum_{x \in E} \pi_x \left (\sum_{y: \,y \neq x} w_{xy}a_{xy}^k f_y \right )^2 \leq \sum_{x \neq y} \pi_x w_{xx}w_{xy}a_{xy}^{2k}f^2_y \\
    & \leq q c^{2k}\sum_{x \neq y} \pi_xw_{xy}f^2_y=q c^{2k}\sum_{y \in E} \pi_y w_{yy} f^2_y \leq q^2c^{2k} \|f\|_\pi^2.\\
\end{split}
\]
In the first inequality we applied Jensen's inequality, while in the last equality we made use of the fact that $\pi$ is invariant for $\W$.

\bigskip 2. With analogous tricks, one obtains the following:
\[\begin{split}
    &\left \|\sum_{x \neq y}a_{xy}^k\J_{xy}\ket{-} \right \|^2_\pi=\sum_{x \in E} \pi_x \left (\sum_{y: \,y \neq x} w_{xy}a_{xy}^k \right )^2\leq \sum_{x \neq y} \pi_x w_{xx} w_{xy}a_{xy}^{2k}\\
    &\leq q c^{2(k-1)} \langle a^2 \rangle_\pi\\
\end{split}
\]
and
\[\begin{split}
    &\left \|\sum_{x \neq y}a_{xy}^k\J^\dagger _{xy}\ket{-} \right \|^2_2=\sum_{x \in E} \pi_x \left (\sum_{y: \,y \neq x} \frac{\pi_y}{\pi_x}w_{yx}a_{yx}^k \right )^2\leq \sum_{x \neq y} \pi_y w_{xx} w_{yx}a_{yx}^{2k}\\
    &\leq q c^{2(k-1)} \langle a^2\rangle_\pi.\\
\end{split}
\]

\bigskip 3. For notation convenience, let us identify $E$ with the ordered set $\{1,\dots, |E|\}$. Let us consider the diagonalization of $\Re(\W)=\mathbb{U}^* \Lambda \mathbb{U}$, where $\mathbb{U}$ is the matrix having as rows the coordinates of a orthonormal basis of eigenvectors (we can pick $\mathbb{U}$ having real entries) and $\Lambda$ is the diagonal matrix of real eigenvalues $\lambda_1=0 \geq \lambda_2=-\varepsilon \geq \dots \geq \lambda_{|E|}$ in decreasing order. Notice that $\Re(\W)_{xx}=-w_{xx}$ and $u_{1x}=\sqrt{\pi_x}$ for every $x \in E$ ($\ket{-}$ the unique eigenvector corresponding to the eigenvalue $0$ and $u_{1x}=\langle \delta_{x}/\sqrt{\pi_x}, - \rangle_\pi=\sqrt{\pi_x}$). Therefore we have
\[-w_{xx}=\Re(\W)_{xx}=(\mathbb{U}^* \Lambda \mathbb{U})_{xx}=\sum_{y=1}^{|E|}\lambda_y u^2_{yx}\leq \lambda_2 \sum_{y \neq 1} u^2_{yx}=\lambda_2 (1-\pi_x),
\]
hence
\[\frac{q}{\varepsilon} \geq 1-\min_{x}\pi_x \geq 1- \frac{1}{|E|} \geq \frac{1}{2}.
\]
\end{proof}
We now return to the series \eqref{eq:series} and \eqref{eq:lambda.k} and focus on $p=1$:
\[\lambda_1^{(k)}=\frac{1}{k!} \left \langle -,\sum_{x \neq y}a_{xy}^k \J_{xy}\ket{-} \right  \rangle_\pi= \frac{1}{k!} \sum_{x \neq y}\pi_x w_{xy}a_{xy}^k,
\]
hence
\[\sum_{k \geq 1} u^k \lambda_p^{(k)}=\sum_{x \neq y}\pi_x w_{xy}(e^{ua_{xy}}-1).\]

Let us consider $k \geq 2$ and $p\geq 2$, then
\[\begin{split}
    \left |\frac{\tr(\B^{(\nu_1)} \S^{(\mu_1)}\cdots \B^{(\nu_p)}\S^{(\mu_p)})}{\nu_1! \cdots \nu_p!} \right |&\leq \frac{\langle a^2 \rangle_\pi}{c}\left (\frac{2q}{\varepsilon} \right )^{p-1}\left (\frac{c}{2}\right )^{k-1} \\
    &\leq
    \frac{q \langle a^2 \rangle_\pi}{\varepsilon}\left (\frac{qc}{\varepsilon} \right )^{k-2}.\\
    \end{split}\]
Above we used that at least 
one of the $\mu_i$'s is equalt to $0$, so that the product under the trace contains a factor $\mathbb{S}^{(0)}= -|-\rangle\langle -|$ and applied the bounds in Lemma \ref{lemma:1}. We further used the fact that $\nu_1!\cdots \nu_p! \geq 2^{k-p}$. 
One can show (\cite[Section 3]{Lezaud1998}) that for $k \geq 3$,
\begin{equation} \label{eq:comb}
\sum_{p=1}^k  \sum_{\substack{\nu_1+\dots +\nu_p=k,\, \nu_i\geq 1\\
\mu_1+\dots +\mu_p=p-1, \, \mu_j\geq 0}} \frac{1}{p}\leq 5^{k-2}
\end{equation}
and we obtain that for $k \geq 2$
\[\left |\sum_{p=2}^k \lambda_p^{(k)} \right | \leq
\frac{q \langle a^2 \rangle_\pi}{\varepsilon}\left (\frac{5qc}{\varepsilon} \right )^{k-2}
\]
Therefore for every $u \geq 0$ such that the R.H.S. of the following inequality is finite, we have
\[\begin{split}
    &|\lambda(u)| \leq \sum_{x \neq y}\pi_x w_{xy}(e^{ua_{xy}}-1)+ \frac{q\langle a^2 \rangle_\pi u^2}{\varepsilon}\sum_{k \geq 0}\left ( \frac{5qcu}{\varepsilon}\right )^{k}\\
    &=\sum_{x \neq y}\pi_x w_{xy}(e^{ua_{xy}}-1)+ \frac{q\langle a^2 \rangle_\pi u^2}{\varepsilon}\left (1-\frac{5qc}{\varepsilon}\right )^{-1}\\
\end{split}\]
Notice that the upper bound of $|\lambda(u)|$ diverges before $u$ exits from the convergence radius of the expression in \eqref{eq:series}: indeed, 
\[\frac{1}{5cq/\varepsilon} \leq \frac{1}{c(2q/\varepsilon+1)} \Leftrightarrow \frac{5q}{\varepsilon} \geq \frac{2q}{\varepsilon}+1 \Leftrightarrow \frac{2q}{\varepsilon} \geq \frac{2}{3}.
\]
In conclusion, we showed that for every $u \geq 0$ for which the R.H.S. is finite, we have
\[\mathbb{E}_\nu[e^{uA(t)}] \leq \left \| \frac{\nu}{\pi} \right \|_\pi {\rm exp} \left ( t\left ( \sum_{x \neq y}\pi_x w_{xy}(e^{ua_{xy}}-1)+\frac{q\langle a^2 \rangle_\pi u^2}{\varepsilon}\left (1-\frac{5cqu}{\varepsilon}\right )^{-1} \right ) \right ).
\]
We can extend the bound to every non-negative $u$ putting the R.H.S. equal to $+\infty$ after it blows up. For getting a bound also for negative $u$, we can repeat the same reasoning for $\tilde{A}(t):=-A(t)$ and we arrive to the expression:
\[\mathbb{E}_\nu[e^{uA(t)}] \leq \left \| \frac{\nu}{\pi} \right \|_\pi {\rm exp} \left (t \left ( \sum_{x \neq y}\pi_x w_{xy}(e^{ua_{xy}}-1)+\frac{q\langle a^2 \rangle_\pi u^2}{\varepsilon}\left (1-\frac{5cq|u|}{\varepsilon} \right )^{-1}\right ) \right )
\]
for every $u \in \mathbb{R}$ (again we extend the R.H.S. beyond the blow up putting it equal to $+\infty$).

\section{Concentration bound}
With a further elementary estimate for the moment generating function, we derive an explicit concentration bound. Notice that, using that $\pi_x w_{xy} \geq 0$, $u >0$ and $c=\max_{x \neq y}|a_{xy}|$, one of the terms appearing in the upper bound can be upper bounded in the following way:
\[\begin{split}
    \sum_{x \neq y} \pi_x w_{xy} (e^{ua_{xy}}-ua_{xy}-1)&=\sum_{x \neq y} \pi_x w_{xy}\sum_{k \geq 2} \frac{(ua_{xy})^k}{k!} \leq \sum_{x \neq y} \pi_x w_{xy}\sum_{k \geq 2} \frac{(uc)^k}{k!} \\
    &=\underbrace{\left (\sum_{x \neq y} \pi_x w_{xy} \right )}_{=\langle k \rangle_\pi} (e^{cu}-cu-1)\\
    \end{split}\]
Hence for every $u\geq 0$ we can write

\[\mathbb{E}_\nu[e^{u(A(t)-t\langle a \rangle_\pi})] \leq \left \| \frac{\nu}{\pi} \right \|_\pi  {\rm exp}\left (t\left ( \langle k \rangle_\pi(e^{cu}-cu-1)+\frac{q\langle a^2 \rangle_\pi u^2}{\varepsilon}\left (1-\frac{5cqu}{\varepsilon}\right )^{-1} \right ) \right ).
\]
Using Chernoff's bound, we obtain that for every $\gamma>0$
\[\mathbb{P}_\nu \left ( \frac{A(t)}{t} \geq  \langle a \rangle_\pi +\gamma \right ) \leq \left \| \frac{\nu}{\pi} \right \|_\pi {\rm exp} \left (-t \underbrace{\sup_{u \in \mathbb{R}} \{\gamma u-h_1(u)-h_2(u)\}}_{=:(h_1+h_2)^*(\gamma)} \right ),
\]
where
\[h_1(u):= \begin{cases} 0 & u <0 \\
\langle k \rangle_\pi (e^{cu}-cu-1) & u \geq 0\end{cases}
\]
and
\[h_2(u)= \begin{cases} 0 & \text{ if } u <0 \\ \frac{q\langle a^2 \rangle_\pi u^2}{\varepsilon}\left (1-\frac{5qcu}{\varepsilon}\right )^{-1} & \text{ if } 0 \leq u < \frac{\varepsilon}{5cq}  \\
+\infty &  \text{ o.w.}\end{cases}
\]
In order to simplify notation, we derive the result for the general function
\[h_2(u)= \begin{cases}  0 & \text{ if } u <0 \\ \frac{u^2}{\zeta-\xi u} & \text{ if } 0 \leq u < \zeta/\xi  \\
+\infty &  \text{ o.w.}\end{cases}
\]
for $\zeta, \xi >0$. It is easy to see that
\[h_1^*(\gamma)= \begin{cases} +\infty & \text{if } \gamma <0\\
\langle k \rangle_\pi g_1 \left ( \frac{\gamma}{\langle k \rangle_\pi c}\right ) & \text{o.w.}\end{cases}, \qquad h_2^*(\gamma)= \begin{cases} +\infty & \text{if } \gamma <0\\
\frac{\zeta \gamma^2}{2}g_2\left ( \xi \gamma\right ) & \text{o.w.}\end{cases},\]
where $g_1(\gamma)=(1+\gamma)\log(1+\gamma)-\gamma \geq \gamma^2/2(1+\gamma/3)$ and $g_2(\gamma)=(1+\gamma/2+\sqrt{\gamma+1})^{-1} \geq (2+\gamma)^{-1}$. We can use Moreau-Rockafellar formula (\cite[Theorem 16.4]{Ro15}) to obtain
\[\begin{split}
    &(h_1+h_2)^*(\gamma)=\inf\{h_1^*(\gamma_1)+h^*_2(\gamma_2):\gamma_1+\gamma_2=\gamma, \, \gamma_1,\gamma_2 \in \mathbb{R}\}\\
    &\geq \inf\left \{\frac{ \gamma_1^2}{2 \left (\langle k \rangle_\pi c^2+ \frac{c \gamma_1}{3} \right )}+\frac{\zeta \gamma_2^2}{2(2+\xi \gamma_2)}:\gamma_1+\gamma_2=\gamma, \, \gamma_1,\gamma_2 \geq 0 \right \} \\.
\end{split}
\]
We can use the fact that $\gamma_1^2/a+\gamma_2^2/d \geq (\gamma_1+\gamma_2)^2/(a+d)$ for non-negative $\gamma_1,\gamma_2$ and positive $a,d$ to obtain
\[\begin{split}
    (h_1+h_2)^*(\gamma) &\geq \inf\left \{\frac{\gamma^2}{2 \left (\langle k \rangle_\pi c^2 + \frac{c \gamma_1}{3}+ \frac{2}{\zeta}+\frac{\xi}{\zeta} \gamma_2 \right )}:\gamma_1+\gamma_2=\gamma, \, \gamma_1,\gamma_2 \geq 0 \right \}\\
    &=\frac{\gamma^2}{2 \left ( \langle k \rangle_\pi c^2 +\frac{2}{\zeta} + \max \left \{\frac{c }{3},\frac{\xi}{\zeta} \right \}\gamma\right )}\\
\end{split}\]
Hence we obtain the following:
\[\mathbb{P}_\nu \left ( \frac{A(t)}{t} \geq \langle a \rangle_\pi +\gamma \right ) \leq \left \| \frac{\nu}{\pi} \right \|_2 {\rm exp} \left (-t\frac{\gamma^2}{2 \left ( \langle k \rangle_\pi c^2+\frac{2c\langle a^2 \rangle_\pi}{\varepsilon} + \frac{5cqu}{\varepsilon}\right )} \right ).
\]

\section{Finite Time and Asymptotic Variance}
For the computations of the bounds on the variance for the generalised currents we derive an expression for the finite and asymptotic time variance for a given current. Applying the equation for the derivative of a matrix exponential to $e^{t\W_u}$:
\[
\frac{de^{t\W_u}}{du}=\int^t_0e^{(t-s)\W_u}\W'_ue^{s\W_u}ds.
\]
We obtain the first moment of $A(t)$:
\[
Z_{\pi,t}^\prime(0)=t \bra{ \pi}\W'_0\ket{-}=t\langle a \rangle_\pi
\]
The second moment can be broken into three parts by applying Leibniz rule when taking the second derivative. The first part comes from the $\W_u'$ in the product:
\[
Z^{\prime\prime (1)}_{\pi,t}(0) = \int_0^t \bra{\pi} e^{(t-s)\W}\W''_0 e^{s\W} \ket{-} ds = t\langle a^2 \rangle_\pi.
\]
The remaining two terms from the $e^{(t-s)\W_u}$ and the $e^{s\W_u}$ are in fact equal, giving:
\[
Z^{\prime\prime (2)}_{\pi,t}(0)=\int_0^t\int_0^{t-s}\bra{\pi}\W'_0e^{r\W}\W'_0 \ket{-} dr ds.
\]
We write $\W'_0\ket{-}=\langle a \rangle_\pi \ket{-}+(\W'_0\ket{-}-\langle a \rangle_\pi \ket{-})$ (recall that $\langle \pi,\W'_0 - \rangle=\langle a \rangle_\pi$) and break the integral into two parts, the first of which:
\[
\langle a \rangle_\pi\int_0^t\int_0^{t-s}\bra{\pi}\W'_0e^{r \W}\ket{-} dr ds = \frac{t^2}{2}\langle a \rangle_\pi^2.
\]

The second part gives:
\[
\begin{split}
\int_0^t\int_0^{t-s}\bra{\pi} \W'_0 e^{r\W}\ket{f} drds=\bra{\pi}\W'_0((e^{t\W}-\mathbb{I})(\W^{-1})^2-t\W^{-1})\ket{-},
\end{split}
\]
where $\ket{f}=\W'_0\ket{-}-\langle a \rangle_\pi \ket{-}$ and $\W^{-1}$ is the pseudoinverse $(\W+\ket{-}\bra{\pi})^{-1}-\ket{-}\bra{\pi}$. Combining all parts gives:
\[
Z''_{\pi,t}(0)=Z^{\prime\prime (1)}_{\pi,t}+2Z^{\prime\prime (2)}_{\pi,t}=t\langle a^2 \rangle_\pi +t^2\langle a \rangle^2_\pi +2\bra{\pi}\W'_0((e^{t\W}-\mathbb{I})(\W^{-1})^2-t\W^{-1})\ket{-}.
\]
Therefore we can write $\sigma^2_\pi(t)$ as
\begin{equation}\label{eq:finitevar}
\sigma^2_\pi(t)=Z''_{\pi,t}(0)- (Z'_{\pi,t}(0))^2=t\langle a^2 \rangle_\pi+2 \bra{\pi} \W_0^\prime ((e^{t\W}-\mathbb{I})(\W^{-1})^2-t\W^{-1})\W_0'\ket{-}
\end{equation}

The asymptotic variance is then obtained by taking the limit:
\[
\sigma^2_{\infty}=\lim_{t\to\infty}\frac{\sigma^2_\pi(t)}{t}=\langle a^2\rangle_\pi-2\bra{\pi}\W^\prime_0\W^{-1}\W^\prime_0 \ket{-},
\]
Which agrees with that obtained from the Central Limit Theorem. We can also recover the static approximate variance by taking the zero time limit:\[
\lim_{t\to0}\frac{\sigma^2_\pi(t)}{t}=\langle a^2 \rangle_\pi
\]

\bibliography{biblio}
\bibliographystyle{abbrv}